\documentclass[10pt,journal,compsoc]{IEEEtran}
\pdfoutput=1
\usepackage{graphicx}
\usepackage{booktabs}   
\usepackage{subcaption} 
\usepackage{color}
\usepackage{xcolor}
\usepackage{algorithm}
\usepackage{algorithmicx}
\usepackage{algpseudocode}
\usepackage{multirow}
\usepackage{amsmath}

\graphicspath{{pic/}{./}}
\definecolor{red}{rgb}{0,0,0}
\definecolor{gray}{rgb}{0,0,0}

\ifCLASSOPTIONcompsoc
  \usepackage[nocompress]{cite}
\else
  \usepackage{cite}
\fi
\ifCLASSINFOpdf
\else
\fi
\begin{document}
%
\title{Benchmarking 50-Photon Gaussian Boson Sampling on the Sunway TaihuLight}
%
%
%
%

\author{Yuxuan Li,
Mingcheng Chen,
Yaojian Chen,
Haitian Lu, \\
Lin Gan,
Chaoyang Lu,
Jianwei Pan,
Haohuan Fu,
and Guangwen Yang
\IEEEcompsocitemizethanks{
\IEEEcompsocthanksitem  Y. Li, Y. Chen, L. Gan, H. Fu, and G. Yang are from Department of Computer Science and Technology, Tsinghua University, Beijing 100084, China.
\IEEEcompsocthanksitem  M. Chen, C. Lu, and J. Pan are from \textsuperscript{1}Hefei National Laboratory for Physical Sciences at Microscale and Department of Modern Physics, University of Science and Technology of China, Hefei 230026, People's Republic of China; and \textsuperscript{2}Shanghai Branch, CAS Center for Excellence in Quantum Information and Quantum Physics, University of Science and Technology of China, Shanghai 201315, People's Republic of China.
\IEEEcompsocthanksitem H. Lu is from the National Supercomputing Center in Wuxi, China.
\IEEEcompsocthanksitem *Corresponding author, Gan Lin (Email address: lingan@tsinghua.edu.cn)}
}
\IEEEtitleabstractindextext{%
\begin{abstract}
Boson sampling is expected to be one of an important milestones that will demonstrate quantum supremacy. 
The present work establishes the benchmarking of Gaussian boson sampling (GBS) with threshold detection based on the Sunway TaihuLight supercomputer. 
To achieve the best performance and provide a competitive scenario for future quantum computing studies, 
the selected simulation algorithm is fully optimized based on a set of innovative approaches, including a parallel scheme and instruction-level optimizing method. 
Furthermore, data precision and instruction scheduling are handled in a sophisticated manner by an adaptive precision optimization scheme and a DAG-based heuristic search algorithm, respectively. 
Based on these methods, a highly efficient and parallel quantum sampling algorithm is designed. 
The largest run enables us to obtain one Torontonian function of a $100\times100$ submatrix from 50-photon GBS within 20 hours in 128-bit precision and 2 days in 256-bit precision.
\end{abstract}

}

\maketitle

\IEEEdisplaynontitleabstractindextext

%
\IEEEpeerreviewmaketitle

\IEEEraisesectionheading{\section{Introduction}\label{sec:introduction}}


\IEEEPARstart{T}{he} extended Church-Turing (ECT) thesis states that a classical computer can efficiently simulate any physical process with only polynomial overheads \cite{supremacyNature}. 
In the 1980s, Feynman observed that many-body quantum problems cannot be efficiently solved by classical computers due to the exponential size of quantum state Hilbert space \cite{feynman1982}. 
This observation inspired Feynman to envision a quantum computer to solve quantum problems.
Efficient quantum algorithms were proposed for classically hard problems, such as integer factoring \cite{factor,book}. 
Today, achieving quantum supremacy based on quantum computers is anticipated to be an important milestone in the post-Moore era. 

Recently, it was found that quantum computers can simulate quantum sampling problems in polynomial time, while classical computers need exponential time unless the polynomial hierarchy (PH) collapses \cite{bsTheory}. 
Therefore, the quantum sampling problem has become a feasible way to demonstrate the quantum supremacy on noisy intermediate-scale quantum (NISQ) devices \cite{nisq}.
According to numerical estimation, quantum sampling with $50 \sim 100$ quantum particles is beyond the computational capabilities of the state-of-the-art supercomputers towards achieving quantum supremacy \cite{howmany}.

Candidates for quantum sampling problems include the instantaneous quantum polynomial time circuit \cite{IQP}, random circuit sampling (RCS) \cite{rcsTheory,rcsTheory2}, boson sampling \cite{bsTheory}, and Gaussian boson sampling (GBS) \cite{gbs,gbsThreshold}. 
For the physical implementation, instantaneous quantum polynomial time circuit and RCS are based on quantum bits, while boson sampling and GBS are based on bosons, such as single photons. 
Since the proposal of GBS greatly simplifies its quantum implementation,
quantum sampling based on bosons is expected to be an important milestone in demonstrating quantum supremacy \cite{supremacyNature,supremacyNPJ}. 

This work establishes the quantum supremacy frontier for the boson-based quantum sampling problem. 
To achieve the best performance and provide a competitive scenario for quantum computers, the selected algorithm is fully optimized based on the Sunway TaihuLight supercomputer, which is one of the most powerful classical computers in the world. 
Our major contributions are as follows:
\begin{itemize}
\item an effective parallel scheme is proposed to reduce the requirements of cache-level storage and obtain an almost ideal load balance of the selected boson sampling algorithm.

\item 
an instruction-based and multiple-precision optimizing scheme that enables customizable data precision modes is designed to guarantee sufficient accuracy of the simulation; an adaptive framework based on upper- and lower-bound estimation is further applied to determine the best precision of the configuration.

\item in terms of instruction-level optimization, a DAG-based heuristic search algorithm is adopted to achieve optimal instruction scheduling for all kernels.
\end{itemize}

\section{Related Works}

In 2019, Google first claimed the quantum supremacy by using 53 superconducting quantum bits in an RCS experiment \cite{googlesupremacy}. 
The quantum device took 200 seconds to sample one instance of a quantum circuit a million times, while classical benchmarking was expected to need 10,000 years on the Summit supercomputer. To confirm and establish this quantum supremacy, there have been many developments in the classical benchmarking algorithm \cite{rcs49,chen201864,rcsParallel,chen2018classical,villalonga2020establishing,rcsTeleport} before and after the experiments. The latest result found by IBM indicates that the calculation can be run on the same supercomputer in less than two and a half days---rather than 10,000 years \cite{pednault2019leveraging,afewdays}.
Furthermore, by leveraging secondary storage of supercomputer, the sampling probabilities of all the $2^{53}$ states can be parallelly computed.
For large number of samples, the average classical calculation time for a single sample will be $2^{53}$ times faster.
Therefore, a more reliable state-of-the-art benchmarking on classical computers is crucial to prove the quantum supremacy. 

Today, many studies have been based on the RCS problem. 
In general, the classical simulators of RCS problems are categorized into three classes.
The first class and the second class are the direct evolutions of the quantum state \cite{de2007massively,de2019massively,smelyanskiy2016qhipster,haner20175,rcs49} and the perturbation of stabilizer circuits \cite{aaronson2004improved,bravyi2016improved,bennink2017unbiased}, respectively.
The tensor network contraction class \cite{chen2018classical,villalonga2018flexible,villalonga2020establishing} is the most suitable method for current flop-oriented architectures such as on the Fugaku \cite{fugaku} and Summit \cite{summit} supercomputers.
In addition, several hybrid algorithms \cite{chen201864,markov2018quantum,pednault2019leveraging} show their great performance in achieving a simulation or benchmark with $\sim 50$ quantum bits.
There are also some efforts \cite{de2019massively, li2019quantum} that attempt to build up the benchmarking of the quantum supremacy based on the RCS problem on the Sunway TaihuLight supercomputer.

The boson sampling problem, which was introduced by Aaronson and Arkhipov \cite{v009a004}, is historically the first protocol to conclusively demonstrate the quantum supremacy.
Several algorithms have been designed and experiments have been conducted referring to such approaches \cite{broome2013photonic,tillmann2013experimental,bs2013science,crespi2013integrated,neville2017classical}. 
However, experimentally, it is actually difficult to scale the boson sampling to high photon numbers owing to its intrinsic photon cost.
GBS \cite{hamilton2017gaussian} and GBS with threshold detection \cite{gbsThreshold} were recently proposed to obtain higher-order photon numbers than those of boson sampling.
There are several classical benchmarks for GBS.
Based on the Titan supercomputer, Brajesh Gupt et al. \cite{gupt2020classical} proposed a benchmark for GBS with threshold detection. However, due to the limitations of the method, it can only handle approximately 22 clicks (or photons) when using the entire Titan supercomputer.

Different from previous works, we establish the first classical benchmarking for GBS with threshold detection based on calculation of the Torontonian function, which has not been used in previous works.
Our design is able to calculate a single Torontonian for 50-photon GBS in less than one day.

\section{Background}

\subsection{Selected Algorithm: Boson Sampling}
In the standard boson sampling experiment, $N$ indistinguishable single photons are sent to an $M$-port linear optical network, and the output scattered photons are detected by $N$ single-photon detectors. The probabilities of these $N$-photon events are related to the permanent function of an $N\times N$ sampling matrix, which is proven to be \#P-hard for classical computers. However, a large-scale experiment suffers from the formidable challenge of preparing $N$ perfect single photons \cite{broome2013photonic,bs2013science,crespi2013integrated,bs2017np,bslossy,bs20,bsScattershot,bsScattershot12}.

\begin{figure}[!ht]
\centering
\includegraphics[width=0.4\textwidth]{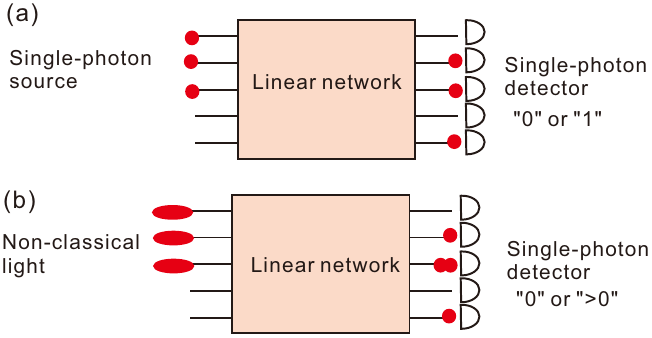}
\caption{(a) Standard boson sampling; (b) GBS with threshold detection.}
\label{fig:boson}
\end{figure}


A practical improvement was proposed to change the input single photons to nonclassical Gaussian light and use single-photon detectors without photon-number resolution to register $N$-click events, which is called GBS with threshold detection \cite{gbs,gbsThreshold,gbsNP,gbsBull}. In this case, the probability of $N$-click events is the Torontonian function of a $2N\times 2N$ sampling matrix, which is expected to still be exponentially hard for classical computers.

In a GBS experiment, before photon detection, the output Gaussian state is described by a covariance matrix $\sum$. The sampling matrix $A$ is then defined as $A=I-\sum^{-1}$, and the probability $p(S)$ of an $N$-click event $S={s_1,s_2,...,s_M}$ is determined by 
\begin{equation}
p(S) = \langle S|\rho_{\sum}|S\rangle = \frac{Tor(A_{\{S\}})}{\sqrt{det(\sum)}}
\end{equation}
, where $\rho_{\sum}$ is the output quantum state of a covariance matrix $\sum$, $S$ is the threshold click pattern with $s_k \in {0,1}$ and $\sum_{k=1}^{M} = N$, and $A_{\{S\}}$is the $2N\times 2N$ submatrix of A by selecting the k-th and $(M+k)$-th row and column when $s_k = 1$. The matrix function $Tor(\cdot)$ is the Torontonian function, which is defined as 
\begin{equation}
\label{equ:tor}
Tor(A) = \sum_{Z\in P_N} (-1)^{N-|Z|}\frac{1}{\sqrt{|det(I-A_Z)|}}
\end{equation}
, where $P_N$ is the power set of ${1,2,...,N}$, and hence, there are $2^N$ determinant terms in the summation \cite{gbsThreshold}. The computational complexity of the Torontonian function is $O(2^N)$, which is the same as the permanent function in standard boson sampling with $N$ photons.

In this work, our major task is to calculate the Torontonian function---which has exponential complexity. 

\subsection{Sunway Architecture}
\label{sec:sunway}

\begin{figure}[ht]
\centering
\includegraphics[width=0.4\textwidth]{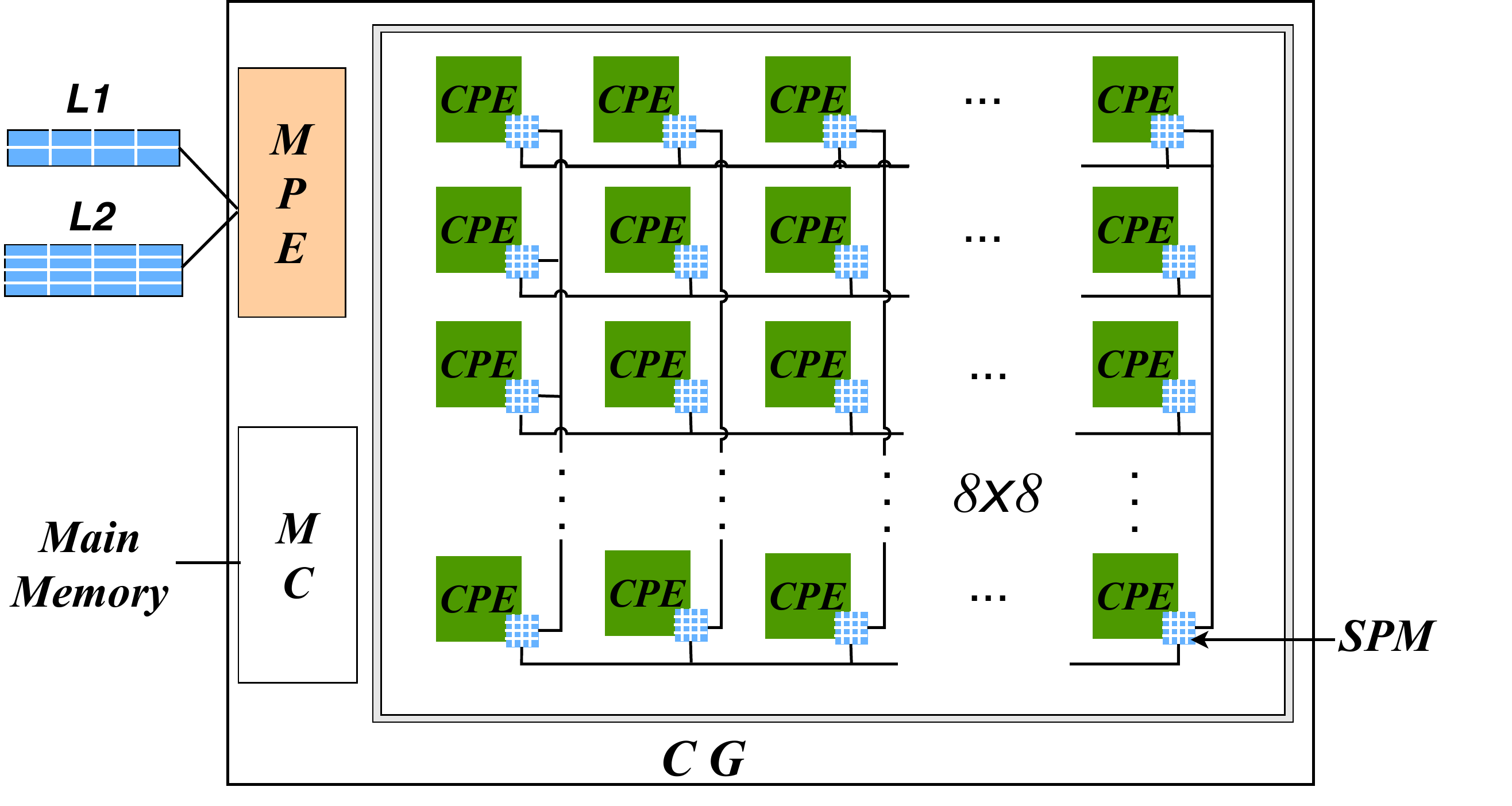}
\caption{Block diagram of a CG.}
\label{fig:sunway}
\end{figure}

To achieve the best simulation performance, 
this work selects the Sunway TaihuLight supercomputer as the classical platform.
The system consists of 40,960 many-core processors called SW26010s and is able to provide 
a peak performance of 125 PFlops and a sustainable performance of 93 PFlops.
Each SW26010 processor consists of four core groups (CGs), which are shown in Figure \ref{fig:sunway}.
Each CG has a memory controller (MC), 
a management processing element (MPE) 
and 64 computing processing elements (CPEs).

As shown in Figure \ref{fig:sunway}, the MPE is a fully functional 64-bit RISC general-purpose core
that contains a 32-KB L1 data cache and 256-KB L2 instruction/data cache.
The CPE is a simplified 64-bit RISC computing-oriented core 
and has a private 16-KB L1 instruction cache and 64-KB scratch-pad memory (SPM).
Direct memory access (DMA) is the major method for the CPEs to access data from global memory.
Furthermore, a thread-level communication is enabled based on register communication.
Thus, inside one CG, the CPE cores within the same row or column are able to communicate with each other directly.

\begin{table}[!t]
\footnotesize
\caption{Main instructions in this work}
\label{tab:ins}
\begin{tabular}{lllllll}
\toprule
Instruction & Description & Latency\\
\midrule
umulqa & 128-bit unsigned & 6 \\
& complete multiplication & \\
uaddo\_carry & 256-bit unsigned addition & 2 \\
usubo\_carry & 256-bit unsigned subtraction & 2 \\
sllow & 256-bit logical left shifting & 2 \\
srlow & 256-bit logical right shifting & 2 \\
vshff & shuffle based on & 1 \\
& two vectors and a mask & \\
vsellt & \textcolor{gray}{vector "less than"} & 1 \\
& conditional selection & \\
\bottomrule
\end{tabular}
\end{table}

A distinctive feature of the Sunway architecture is its 256-bit large integer instruction sets,
as shown in Table \ref{tab:ins}.
The first five instructions (multiplication, addition, subtraction and shifting) belong to the 256-bit large integer instruction sets.
However, the multiplication instruction is not a straightforward 256-bit version; instead, it performs an unsigned multiplication of two 128-bit unsigned numbers and obtains a full 256-bit result.

\begin{figure}[!ht]
\centering
\begin{subfigure}{.23\textwidth}
\includegraphics[width=1\textwidth]{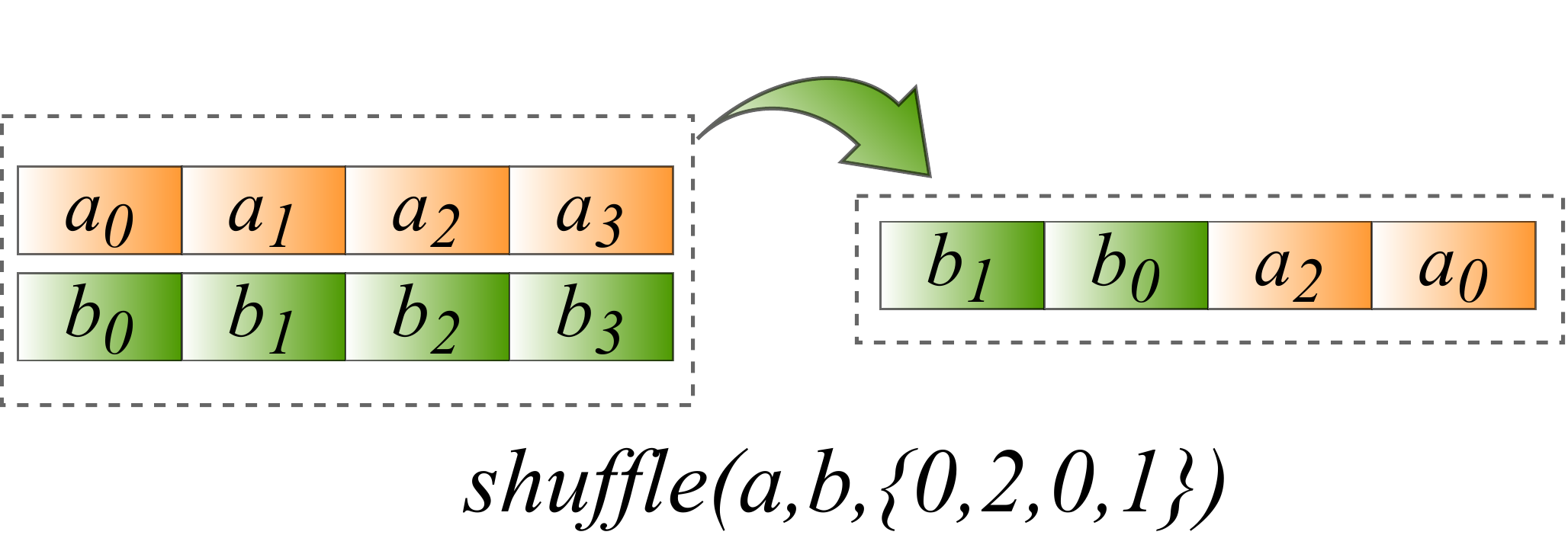}
\caption{}
\label{fig:vshff}
\end{subfigure}
\begin{subfigure}{.23\textwidth}
\includegraphics[width=1\textwidth]{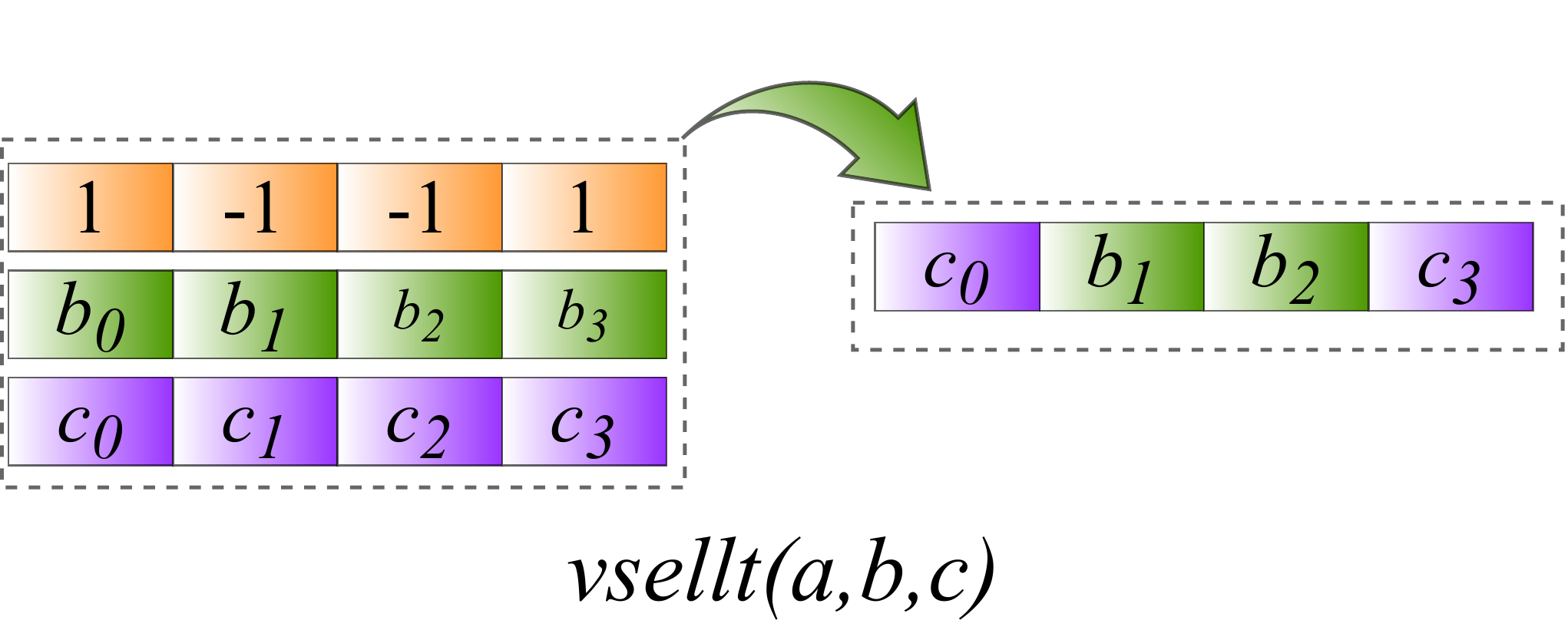}
\caption{}
\label{fig:vsellt}
\end{subfigure}
\caption{(a) Example of vshff instruction; (b) example of vsellt instruction}
\label{fig:sunwayins}
\end{figure}

The rest of the instructions in the table are regular vector instructions and are useful in optimizing the boson sampling process.
The parameters of instruction \textit{vshff} is $a$, $b$ and \textit{mask} are shown in Figure \ref{fig:vshff}.
$a$ and $b$ are two 256-bit registers that contain four 64-bit numbers.
\textit{mask} provides information on how to construct a new register from $a$ (the first two digits of \textit{mask}) and $b$ (the last two digits of \textit{mask}).
Figure \ref{fig:vshff} shows an example: based on the four digits of \textit{mask}, 0 and 2 of A (corresponding value a0 and a2), and position 0 and 1 of B (corresponding value b0 and b1) are selected to construct a new register.

The conditional selection instruction \textit{vsellt} takes three parameters $a$, $b$, and $c$, which are all 256-bit registers composed of four 64-bit numbers.
$a$ is the conditional register that chooses between $b$ and $c$.
As shown in Figure \ref{fig:vsellt}, if one of the 64-bit numbers in $a$ is less than 0, then the corresponding 64-bit number of the result comes from the corresponding bits of $b$ and vice versa.

\section{Parallel Scheme Design}
This section introduces the parallel scheme of solving Equation \ref{equ:tor}, 
including partition and storage strategies for load balancing and the LDM capacity issues.

\subsection{Partition Strategy}
\label{subsec:partition}

For Equation \ref{equ:tor}, 
\textcolor{gray}{the best dimension to partition is the enumeration of $Z$ due to its embarrassing parallelism.}
For convenience, we map each $Z$ to \textcolor{gray}{a number mask "$mask_{z}$"} ranging from $1$ to $2^N-1$.
If $i \in Z$, the i-th bit in the binary representation of $mask_{z}$ is "$1$" and vice versa.

A naive partition strategy can be directly obtained by evenly dividing the $1 \sim 2^N-1$ mask into every CPE from different CGs. 
However, 
while two continuous numbers must have different numbers of "one" in the binary representation,
the matrix size and computing load of each CPE are different.
Therefore, the simple and naive strategy brings a severe thread-level load imbalance.

\begin{algorithm} 
\footnotesize
    \caption{Calculate $Torontonian(A)$}  
    \label{alg:tor}
    \begin{algorithmic}[1]
        \Require Matrix $A$
        \Ensure $A$ is Hermitian positive definite
        \Function {$Torontonian$}{$A$}  
            \State $result \gets 0$
            \For {$i_{|Z|} = 1 \rightarrow N$}
                \State $mask_z \gets \Call{get\_kth\_mask}{N, i_{|Z|}, \frac{\tbinom{N}{i_{|Z|}} * rank}{nproc}}$
                \State $maskEnd \gets \Call{get\_kth\_mask}{N, i_{|Z|}, \frac{\tbinom{N}{i_{|Z|}} * (rank+1)}{nproc}}$
                \While {$mask_z \neq maskEnd$}
                    \State $A_Z \gets \Call{gen\_$A_Z$}{mask_z}$
                    \State $det \gets \Call{get\_determinant}{I-A_Z}$
                    \State $result \gets result + (-1)^{N-i_{|Z|}}\frac{1}{\sqrt{|det|}}$
                    \State $mask_z \gets \Call{get\_next\_mask}{mask_z}$
                \EndWhile  
            \EndFor
            \State \Return{$result$}  
        \EndFunction  
    \end{algorithmic}  
\end{algorithm}  

To avoid this situation, we design a new partition strategy with high load balancing.
We first divide the workloads of matrices into $N$ parts according to the matrix size $|Z|$.
In each part, we divide all masks evenly and continuously into every \textcolor{gray}{process}.
Algorithm \ref{alg:tor} describes the Torontonian function with the new partition strategy.
Line 3 indicates the enumerating of $N$ parts;
in Line 4 and Line 5, the masks assigned to a process are determined;
in Line 7, the calculating matrix $A_Z$ is generated by the current mask;
Line 8 is the determinant calculation that is the most expensive part;
Line 9 is the update of the result; and
in Line 10, the next mask is derived from the current mask, and then, the next loop begins.
In a certain part, each CPE processes a matrix of the same size, obtaining the best load balance.

\begin{algorithm} 
\footnotesize
    \caption{Get Matrix Mask}  
    \label{alg:mask}
    \begin{algorithmic}[1] 
        \Function {get\_kth\_mask}{$N, |Z|, k$}  
            \State $l \gets -1$
            \State $mask_z \gets 0$
            \For{$i = 0 \rightarrow |Z|-1$}
                \For{$j = l+1 \rightarrow N-|Z|+i+1$}
                    \State $cnt \gets \Call{C}{$N-j-1$,$|Z|-i-1$}$
                    \If{$cnt > k$}
                        \State $l \gets j$
                        \State $mask_z \gets mask_z \ or \ 2^{N-j-1}$
                    \EndIf
                    \State $k \gets k - cnt$
                \EndFor
            \EndFor
            \State \Return{$mask_z$}
        \EndFunction  
        \Function {get\_next\_mask}{$mask$}  
            \State $cto \gets \Call{count\_tail\_one}{$mask$}$
            \State $x \gets mask - (2^{cto}-1)$
            \State $ctz \gets \Call{count\_tail\_zero}{$x$}$
            \State $mask_{next} \gets x - 2^{ctz - cto - 1}$
            \State \Return{$mask_{next}$}
        \EndFunction  
        \State  
    \end{algorithmic}  
\end{algorithm}  



In the naive strategy, we only need to calculate $mask_z = 2^N * rank / nproc$ to obtain the first mask of a certain process (or obtain the k-th mask) and to calculate $mask_z + 1$ to obtain the next mask. 
$nproc$ denotes the total number of processes, and $rank$ denotes the rank of \textcolor{gray}{a certain process}.

\textcolor{gray}{
However, the two functions for obtaining the mask, the pseudocodes of which are shown in Algorithm \ref{alg:mask}, are rather complicated in the new strategy.
The \textit{get\_kth\_mask} function is based on a bit-by-bit algorithm that determines the position of every "1" from high to low in binary representation.
In each round of the outer loop, the position of a "1" is determined.
In the inner loop, we try each possible position one by one according to the comparison results of the combinatorial enumeration "\textit{cnt}" and the current "$k$".
If \textit{cnt} is greater than "$k$", then we set the position to "1" and begin the next round of the outer loop; otherwise, we subtract \textit{cnt} from "$k$".
}
\textcolor{gray}{
The main task of the \textit{get\_next\_mask} function is to move the consecutive "1"s at the tail of the binary representation to a proper position.
For instance, consider that the current mask is "010011".
To obtain the next mask "001110", we need to move the last two "1"s to the fourth position (beginning with 1) and move the first "1" to the next position.
Therefore, a bit operation algorithm is designed.
$cto$ and $ctz$ represent the number of consecutive ones and zeros at the tail, respectively.
$cto$ is used to count the consecutive "1"s at the tail, while $ctz$ is used to indicate the proper position.
}

Such a partition scheme is also applicable to thread-level parallelisms with trivial modifications.
\subsection{Storage Strategy}
\label{subsec:storage}
Neither the input nor the calculating matrix can be fit to the 64 KB LDM space if the scale $N$ is too large.
For instance, when $N$ is 45 based on 256-bit multiple precision, the storage requirement of a matrix is $(2N) \times (2N) \times \textit{sizeof(complex)} = 90 * 90 * 64 = 506.25~KB$, and \textcolor{gray}{far} exceeds the LDM capacity.
In addition, when reserving some LDM space to store the input matrix to avoid extra DMA copies, the insufficiency of capacity can be even worse.
Therefore, we propose a storage strategy to lower the LDM requirement of both the input matrix $A$ and calculating matrix $A_Z$.


For $A$, we use lossy compression and a matrix symmetric. As scientific measurement data, there are only three to five significant digits in each element of matrix $A$. 
Therefore, \textcolor{gray}{a 16-bit number (with an error of $\pm1.525\times10^{-5}$) is sufficient to describe the measurement data}. Additionally, in our implementation, 32-bit lossy compression is supported for higher-precision requirements.

Another reason that brings storage improvements is the usage of a matrix symmetric.
Owing to the physical properties of the matrix $A$, 
if we represent matrix $A$ in the form of a block matrix with four blocks as
\begin{equation}
\label{equ:blockA}
A = 
\begin{bmatrix}
A_{00} & A_{01} \\
A_{10}& A_{11}
\end{bmatrix}
\end{equation}
then the following three symmetries are satisfied: 
\begin{enumerate}
\item $A$ is Hermitian
\item $A_{01}$ and $A_{10}$ is symmetric
\item $A_{00} = \overline{A_{11}}$
\end{enumerate}
Now, only half of $A_{00}$ and $A_{01}$ need to be stored, and thus, only approximately $1/4$ of the LDM space is required.


\begin{figure}[!ht]
\centering
\includegraphics[width=0.35\textwidth]{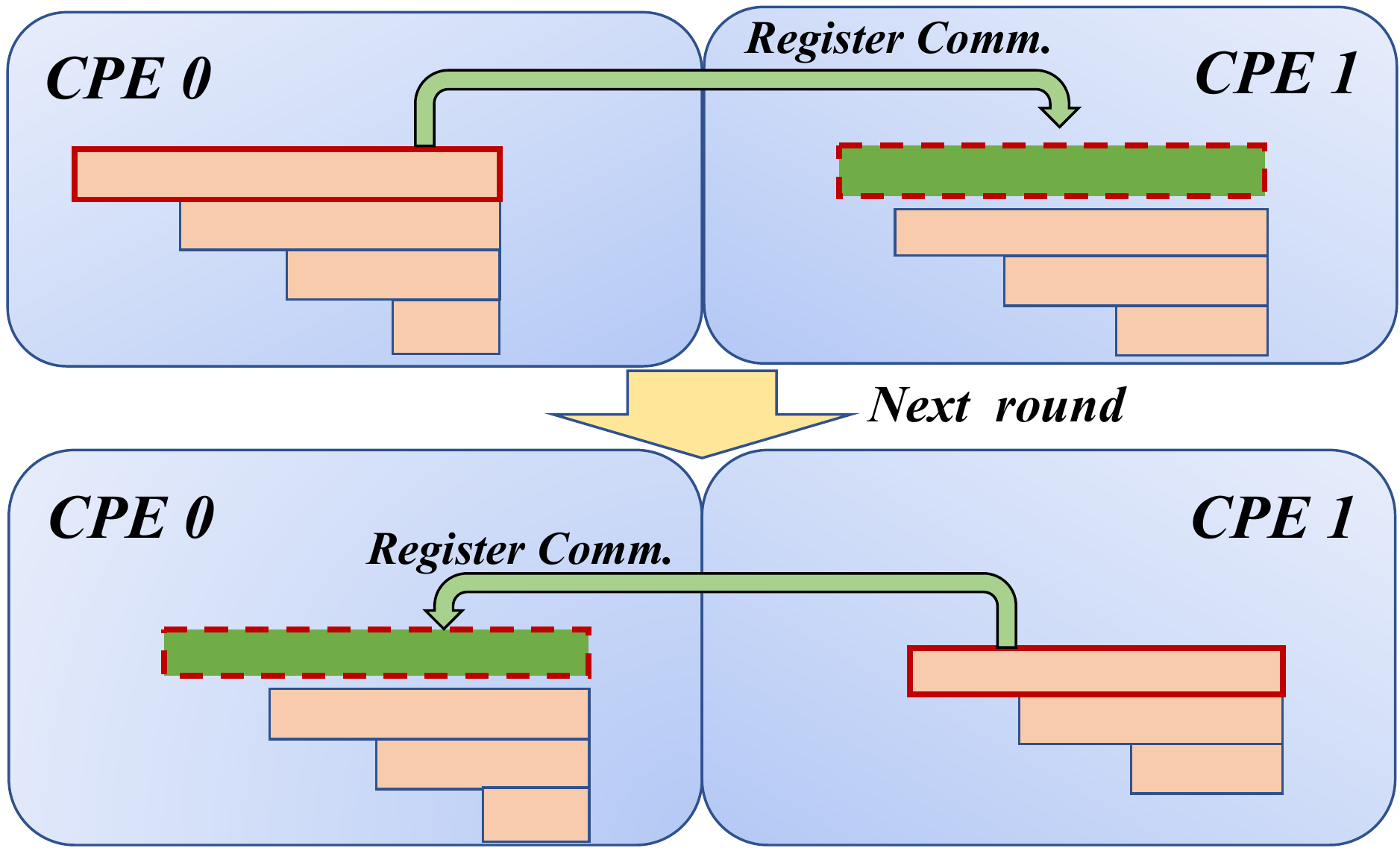}
\caption{Gaussian Elimination based on register communication. 
The upper panel and lower panel represent two adjacent rounds of reduction.
Each block indicates a row of matrix.
Since the matrix is an upper triangular matrix, the lower block has fewer elements than the upper block.
The block with a red box indicates the row where the pivot is located.
The green arrow indicates the direction of register communication.
The green block indicates the buffer for receiving the row with pivot from the other CPE.}
\label{fig:gauss}
\end{figure}

For calculating the matrix $A_Z$, 
we use the Hermitian symmetric and distribute the matrix into several CPEs to save the LDM space.
Since $A$ is Hermitian positive definite, it is not difficult to prove that $I-A_Z$ is also Hermitian positive definite, and in each round of row reduction, the matrix is always Hermitian. 
Thus, we can save half the space of the matrix when performing calculations, 
i.e., the upper triangular part of the matrix, which saves half of the LDM space.

It is still not enough to put the whole calculating matrix into LDM.
The matrix must be scattered into different CPEs, which brings an extra cost when a CPE needs data from other CPEs.
Fortunately, the register communication mechanism of the Sunway architecture provides an effective and efficient way to exchange data among different CPEs.

Figure \ref{fig:gauss} illustrates our distribution strategy and the usage of register communication when the matrix is scattered to two CPEs.
Each row of the matrix is alternately assigned to two CPEs.
In the upper panel, the row with pivot is located at CPE 0.
In this round, CPE 0 sends the row with pivot to CPE 1 via register communication.
Then, each CPE multiplies this row by a proper scalar and adds the scaled row to other rows to \textcolor{gray}{finish} reduction.
In the next round of row reduction, as shown in the lower panel, the row with pivot comes to CPE 1, and thus, the data need to be transferred from CPE 1 to CPE 0.
The process of reduction is the same as in the previous round.

Owing to the two strategies for the calculating matrix, the LDM is now enough to store all the required data.
When $N$ is 45 and the matrix is scattered to eight CPEs, each CPE is occupied $506.25~KB / 2 / 8 = 31.64~KB$ LDM space, 
where $506.25~KB$ is mentioned in the beginning of this subsection, the division factor $2$ is due to Hermitian symmetry, and $8$ is due to the utilization of eight CPEs.
In addition, when using 32-bit compression, the size of the input matrix is $(2N)*(2N)*2*4 / 4 = 16.17~KB$.
Therefore, the total LDM requirement is approximately $31.64~KB + 16.17~KB = 47.81~KB$, which is less than the capacity limit of $64~KB$.

\section{Adaptive Multiple-Precision Design}
\subsection{Motivation}
\label{subsec:prec-motivation}
For real-world data, the result of the Torontonian function is a very small decimal, which results in a high-precision requirement. 
For instance, when $N$ is 45, the result is $1.44 \times 10^{-25}$.
According to Equation \ref{equ:tor}, the result $Tor(A)$ is accumulated from the summation of massive reciprocal square roots of the determinant.
We can prove that the value of each reciprocal square root is greater than 1. 
Thus, at least 25 decimal digits (roughly equivalent to 83 binary bits) is required to achieve adequate accuracy in the determinant calculation.
According to the IEEE 754 standard \cite{ieee754}, 
the double-precision type, which has only 53 binary bits of significant precision, 
is not competent to represent these numbers.
Therefore, a customized multiple-precision type must be designed for the Torontonian function.
To design such a precision type, we first consider the upper and lower bound analysis of the absolute value of intermediate results.
The actual upper bound depends on the summation order of the Torontonian function.
If all positive numbers are summed first and all negative numbers are added afterward, 
then the intermediate results will be very large and proportional to the times of summation (i.e., $O(2^N)$).
However, the result of the Torontonian function is guaranteed to be a very small number.
Thus, the overflow of the summation can be omitted.
In other words, the part of the intermediate results greater than 1 can be \textcolor{gray}{truncated}.

The upper bound, which cannot be omitted, is reached when calculating the reciprocal of pivot in the Gaussian elimination. The value of pivot is difficult to obtain, but fortunately, it can be proved to be smaller than its corresponding
determinant, so we can replace all of them with the value of the determinant.
Furthermore, it can be shown that the determinant of $I-A$ is less than any of its principal minors, 
which means that the upper bound can be estimated by the reciprocal of the determinant of $I-A$. To make an estimation without precise computation, $I-A$ can be calculated with this approximation with $|Z|=N$:
\begin{equation}
\label{equ:upbound}
det(I-A_Z) \approx ((1-a) - \frac{k}{1-a}|Z|)^{|Z|} \cdot (1-a)^{|Z|}
\end{equation}
, where $k$ is a correction factor depending on matrix $A$ that has an observed range of $0.0009\sim0.0035$, 
and $a$ is the representative value of the diagonal element $a_{ii}$. 
In real-world experiments, the range of $a$ is $0.16 \pm 0.06$. 
In addition, the value of the determinant when $N=60$ will reach $10^{-6}$, 
which means that the upper bound can be estimated to be $10^7$.

The lower bound is reached at the final result of the Torontonian function. 
Equation \ref{equ:lowbound} is an approximation of the final result based on the theorem of matrix analysis and the result of $det(I-A)$:
\begin{equation}
\label{equ:lowbound}
Tor(A) \approx (((1-a)-\frac{kN}{1-a})^{-1}-1)^N
\end{equation}
, where $k$ and $a$ are the same as in Equation \ref{equ:upbound}. 
When $k$, $a$, and $N$ are not too large ($k<0.0035$, $a<0.2$, $N<60$), we have $kN/(1-a)\ll1-a$.
Therefore, the approximation is close to ${(1/(1-a)-1)}^N$, 
which can be guaranteed to be smaller than the actual result. 
Then, our approximation can be trusted to be a lower bound of the actual result. 
With different $k$ and $a$, there will be some variances in the value of the Torontonian function. 
Taking the range of these parameters into consideration, the lower bound can reach $10^{-61}$ at $N = 60$.

According to the bound analysis, the range of intermediate results falls into a range of $10^{-61}$ to $10^7$, which can be represented by a large fixed-point number. 
Therefore, we choose a fixed-point number as our precision type, which is much faster than a floating-point number.

\subsection{Multiple-Precision Design}
The large integer instruction set in the Sunway architecture is very suitable for implementing fixed-point numbers.
As mentioned in Section \ref{sec:sunway}, 
the Sunway processor supports 256-bit large integer operations, 
which implies that up to 256-bit fixed-point calculations can be implemented by hardware instructions directly or almost directly.
According to the bound analysis, 256-bit fixed-point precision is sufficient for up to $N=60$, which is an impossible task for a classical computer.

However, it must be noted that there is no 256-bit large integer multiplication in the architecture. 
The largest multiplication instruction is the 128-bit unsigned multiplication.
Constructing the 256-bit signed multiplication requires numerous instructions based on the supported 128-bit multiplication.
It is too expensive to perform the 256-bit signed multiplication.

To decrease the cost of multiplication, we propose an adjustable multiple-precision design.
In addition to the 256-bit precision, we mainly provide a precision mode of 128-bit fixed-point numbers, which is much less expensive in terms of multiplication.
The 128-bit precision mode is not competent for all cases but is much faster than the 256-bit precision mode when it works.

For the scaling factor, we just use a global factor for every fixed-point number. 
Unifying the scaling factor is beneficial to the performance efficiency.



\subsection{Operation Design}
\label{subsec:operation}

In the determinant solver of the Torontonian function, $O(2^NN^3)$ times of multiplications are required, which is the bottleneck of the whole calculation.
Constructing signed multiplication based on the unsigned multiplication is the crucial challenge in multiple-precision design.
A simple method is to convert negative numbers to positive numbers by using if statements.
However, the if statements may result in a failure of branch predictions, which is a hazard to performance.
Thus, avoiding if statements is considered the best way to enhance performance.

Take a 128-bit multiplication between 128-bit signed numbers $a$ and $b$ as an example. For convenience, we assume that $a$ and $b$ are both negative.
The complements of $a$ and $b$ are expressed as $(2^{128} - |a|)$ and $(2^{128} - |b|)$. The unsigned multiplication is represented below: 

\begin{equation}
\label{equ:mul}
(2^{128} - |a|) * (2^{128} - |b|) = 2^{256} - (|a| + |b|) * 2^{128} + |a| * |b|
\end{equation}

Therefore, the 256-bit result produced by unsigned multiplication is $|a| * |b| - (|a| + |b|) * 2^{128}$, in which the term of $2^{256}$ is \textcolor{gray}{truncated}.
The result is not equal to the correct one, which is $|a| * |b|$. 
It worth noting that if $a$ and $b$ are 256-bit signed numbers, then the second term of Equation \ref{equ:mul} is $(|a| + |b|) * 2^{256}$, which can be truncated.
\textcolor{gray}{This indicates that the 256-bit results of the signed multiplication can be taken from the unsigned multiplication between two 256-bit signed numbers.}

However, as mentioned above, there is no 256-bit multiplication instruction in the Sunway architecture.
Instead, three times of 128-bit multiplication are required, i.e., $a * b = a_{[128, 256)} * b_{[0, 128)} + a_{[0, 128)} * b_{[128, 256)} +  a_{[0, 128)} * b_{[0, 128)}$.
Because $a$ and $b$ are essentially 128-bit numbers, $a_{[128, 256)}$ and $b_{[128, 256)}$ are either $0$ or $-1$; i.e., $a_{[128, 256)} * b_{[0, 128)}$ is equal to $0$ or $-b_{[0, 128)}$.
Therefore, there is no need to actually perform multiplication for $a_{[128, 256)} * b_{[0, 128)}$ and $a_{[0, 128)} * b_{[128, 256)}$.
In contrast, vectorized conditional selection instructions are competent.

\begin{algorithm}  
    \caption{128-bit signed multiplication}  
    \label{alg:mul}
    \begin{algorithmic}[1]
        \Require 128-bit signed number $a$ and $b$
        \Function {mul128}{$a$, $b$}  
            \State $c \gets \Call{umulqa}{a, b}$
            \State $a_{sign} \gets \Call{vshff}{0, a, 0x55}$ \Comment{\Call{shuffle}{0, a, \{1, 1, 1, 1\}}}
            \State $b_{sign} \gets \Call{vshff}{0, b, 0x55}$ \Comment{\Call{shuffle}{0, b, \{1, 1, 1, 1\}}}
            \State $a_{128} \gets \Call{vsellt}{b_{sign}, a, 0}$
            \State $b_{128} \gets \Call{vsellt}{a_{sign}, b, 0}$
            \State $c \gets (c - a_{128} - b_{128}) >> \textit{scaling\_factor}$
            \State \Return c
        \EndFunction  
    \end{algorithmic}  
\end{algorithm}

Algorithm \ref{alg:mul} describes our 128-bit signed multiplication design.
The first line describes 128-bit unsigned multiplication.
The second and third lines are prepared for conditional selection instructions.
By \textit{vshff} instruction, the sign bit of $a$ is extracted and put into \textcolor{gray}{high 128 bits} of $a_{sign}$. The low 128 bits of $a_{sign}$ is $0$.
In the next line, according to $a_{sign}$, the vectorized conditional selection instruction \textit{vsellt} is applied to make a choice between $b$ and $0$.
Last, $c$ is calculated and scaled in Line 7.

Therefore, in our method, a total of eight instructions are adequate for 128-bit signed multiplication. 
The 256-bit signed multiplication is much more difficult but shares the same idea.
In our implementation, 19 instructions are sufficient.

Division, which is used for calculating the scalar applied in the elimination, is the second expensive operation.
Although only $O(2^NN^2)$ times of divisions are required, the time complexity of division is much higher than that of multiplication.
In our work, we use $O(2^NN)$ times of reciprocal calculations and $O(2^NN^2)$ times of multiplications, instead of using division directly.
In general, the \textcolor{gray}{reciprocal optimization} results in a decrease in precision;
however, \textcolor{gray}{our multiple precision design is enough to withstand the loss of accuracy.}
We choose trial division to implement the reciprocal.
The time complexity is $O(b)$, where $b$ represents the bit length of the multiple-precision fixed-point number.

The last operation required is the reciprocal square root.
It has the same time complexity with division but is only called $O(2^N)$ times, and thus, it is \textcolor{gray}{evidently} not a bottleneck in the Torontonian function.
Similar to division, a bitwise method is applied to the reciprocal square root.

\subsection{Adaptive Precision Strategy}
The final task of our precision design is to determine the scaling factor and bit length of the multiple-precision type (i.e., 128 bits or 256 bits).
If the scaling factor is too small or the bit length is too short, 
then the results may have an overflow or insufficient significant digits.
In contrast, if the bit length is too long, then the performance efficiency will be affected.
In this subsection, we propose a strategy to select a proper scaling factor and bit length automatically.

The scaling factor depends on the upper bound of all intermediate results.
As explained in Subsection \ref{subsec:prec-motivation}, 
the upper bound can be represented by the reciprocal of determinant of $I-A$.
As long as we calculate the determinant of $I-A$ in advance, the upper bound can be obtained, hence the scaling factor.
The choice of bit length is much more complicated and depends on several factors including both the upper bound and the lower bound.
In our adaptive precision strategy, Equation \ref{equ:lowbound} is chosen to approximate the trusted lower bound.

In addition, bits owing to the accumulated error should be taken into account.
Since each element has $N^2$ times of elementary arithmetic in a determinant calculation, we assume that one time of determinant calculation will produce an error of $\alpha N^2$, where $\alpha$ is a correction factor.
Meanwhile, $2^N$ times of determinant calculations are required.
Thus, we use $\alpha 2^N N^2$ to estimate the accumulated error, where $\alpha$ is taken as \textcolor{red}{0.5} by some experiments.

Last but not least, the significant digits required (usually for physical experiments, this is three decimal digits) should be considered.

The final bit length can be represented as follows:
\begin{equation}
\label{equ:bitlen}
B = B_{lower} + B_{upper} + B_{sgn} + B_{accum}
\end{equation}
, where $B$ denotes the total bit length required, $B_{lower}$ denotes the part of the bit length derived from the upper bound (or scaling factor),
$B_{upper}$ denotes the part derived from the upper bound,
$B_{sgn}$ denotes the part derived from the significant digit requirement, and
$B_{accum}$ denotes the part derived from the accumulated error.
According to the relationship between $B$ and 128, the 128-bit precision mode or 256-bit precision mode is chosen.

\section{Optimal Instruction Scheduling}
\subsection{Framework}
To obtain the optimal instruction scheduling, the corresponding hardware features must be considered carefully.
For each CPE from the Sunway processor, 
there are two hardware pipelines with different functions.
The first pipeline mainly takes charge of floating-point or vector instructions, 
and the second pipeline mainly takes charge of memory access and branch instructions. 
In addition, integer scalar instructions are processed by both pipelines.
In our multiple-precision design, most instructions are large integer instructions, which belong to vector instructions and are thereby taken charge of by the first pipeline.
Only about $15\%$ of the instructions belong to the second pipeline in each kernel.

The performance (or the pipeline stall) of the first pipeline determines the overall performance.
Two hazards affect the pipeline efficiency in the Sunway architecture.
One is the read-after-write (RAW) data hazard; the other is the writeback stage structural hazard, which conflicts with the writeback of general-purpose registers.
Given both hazards, it is hard to find an optimal instruction scheduling directly.
Thus, we turn to optimization methods to find the optimal instruction scheduling.

For the instructions assigned to the second pipeline, we manually insert them into the scheduled sequence of instructions that belong to the first pipeline.
To eliminate computation dependence and avoid data hazards, \textcolor{gray}{software pipeline technology is applied}.
As a result, these inserted instructions do not increase the number of total clocks in our case.
In other words, all time cost comes from the first pipeline.

Consequently, our instruction scheduling framework is as follows:
\begin{description}
\item{\textbf{Stage 1}} Distinguish which pipeline to which each instruction belongs.
\item{\textbf{Stage 2}} Search an optimal scheduling from instructions of the first pipeline.
\item{\textbf{Stage 3}} Manually insert instructions of the second pipeline into the scheduled instruction sequence.
\end{description}

\subsection{Heuristic Search Based on DAG Constraints}
In our instruction scheduling framework, Stage 2 is the critical stage that determines the final performance.
A heuristic search method is proposed to address this stage.

We choose the A-Star search algorithm \cite{wikiastar} as the framework. 
To search in heuristic order, the A-Star algorithm selects the next state that minimizes
$f(n) = g(n) + h(n)$, where $n$ is the next state be selected, $g(n)$ is the cost of the selected state set,
and $h(n)$ is a heuristic function that estimates the lowest cost from the current state set to the goal.
In our work, $g(n)$ is the execution clocks of all selected instructions, 
and $h(n)$ is the number of remaining instructions.

However, only an A-Star search is insufficient with regard to the instruction scheduling with a scale of more than 30 instructions.
More constraints to prune are necessary.
Using a directed acyclic graph (DAG) to represent constraints is very suitable to the instruction scheduling. 
In general, 
a node in a DAG represents an instruction; an edge $a \rightarrow b$ in a DAG indicates that $b$ has data dependence on $a$.
With regard to the heuristic methods based on DAGs, a previous work \cite{dagheuristic} surveyed three DAG construction algorithms and twenty-six proposed heuristics methods and evaluated their performance exhaustively.

\begin{figure}
\centering
\includegraphics[width=0.51\textwidth]{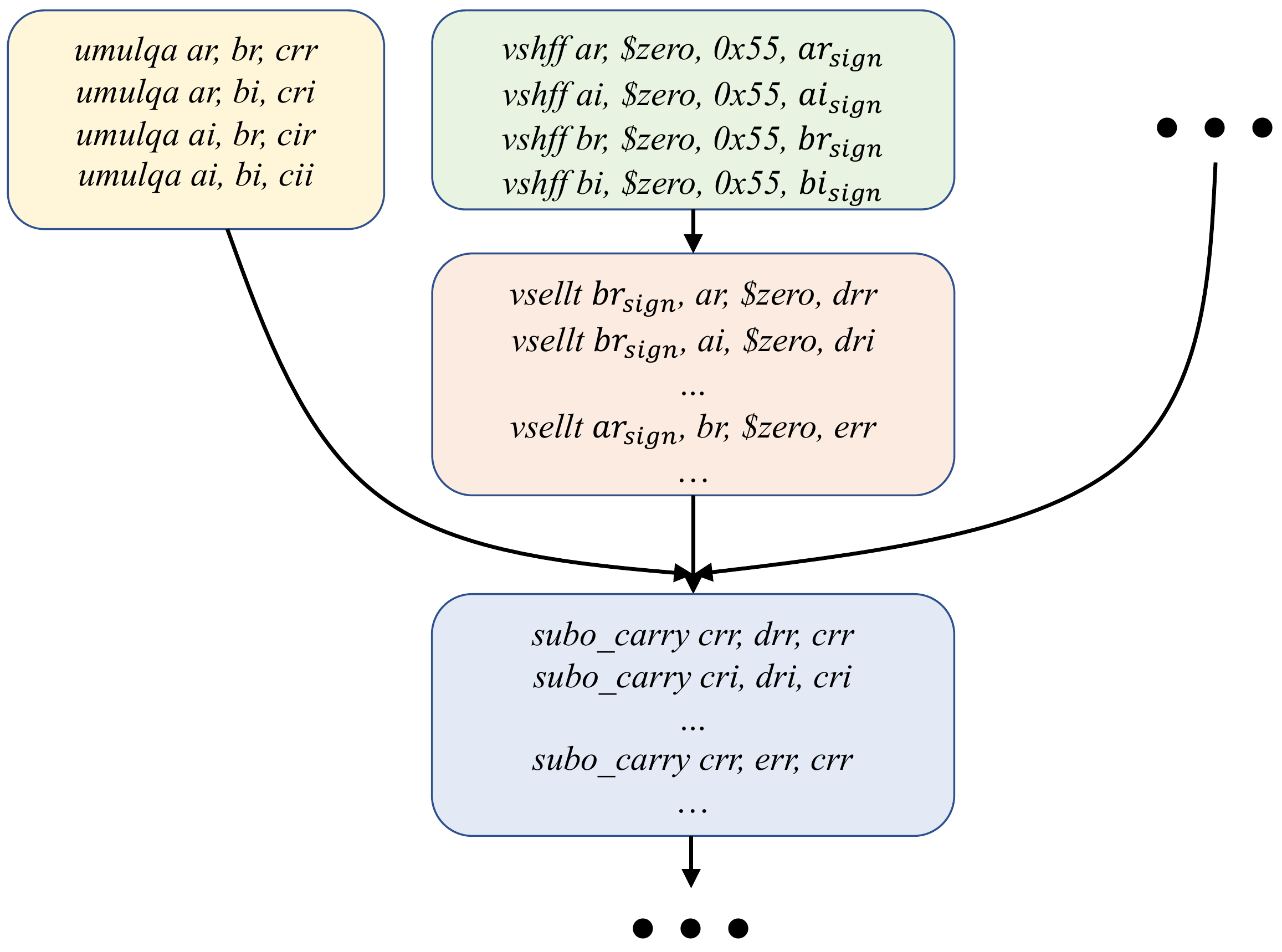}
\caption{Part of the instruction DAG for the 128-bit reduction of Gaussian elimination.}
\label{fig:dag}
\end{figure}

Different from existing methods, a node in a DAG represents a sequence of instructions in our method. 
Due to the properties of complex operations, instructions in the kernel are symmetrical to some extent.
For example, computing of the real and imaginary parts is symmetrical.
Moreover, if the operation is fused multiply-add (FMA), then
the instructions can be divided into 4 symmetrical parts.
For each symmetrical part, the corresponding instructions are the same instructions with different operands.
We assign the corresponding instructions from different parts into the same DAG nodes.
Then, we can set an order for these corresponding instructions distinguished by the operands, and all nodes share the same order. 
Setting such an order does not lead to a loss of the optimal solution because
(1) swapping corresponding instructions in some DAG nodes cannot affect the structural hazard in the writeback stage since all corresponding instructions have the same clocks, and
(2) data hazards cannot be reduced by swapping since the best choice is for all DAG nodes to share the same order.

Figure \ref{fig:dag} shows the main part of the instruction DAG of a 128-bit reduction.
In the top left yellow panel, the order within the node is determined as "$crr$", "$cri$", "$cir$", and "$cii$", which are four different combinations of the real and imaginary parts of two complex numbers.
The remaining DAG nodes all maintain the same internal order.
In addition, in the middle orange panel, we can put eight instructions into this node due to the symmetry between "$b, ar_{sign}$" and "$a, br_{sign}$" rather than the complex symmetry.
Through the optimization, the efficiency of the heuristic search is increased many times, and thus, we can obtain the optimal results.

Table \ref{tab:inssche} lists the clocks of different methods in different computing parts. 
"128 Outer" represents the clocks of the instruction block in the outer loop, which is called $O(N^2)$ times (while the inner loop (or the reduction of Gaussian elimination) is called $O(N^3)$ times) when using 128-bit precision mode.
For the three different methods, "Origin" represents results without any scheduling; "Manual" represents results by a redundant removal and a relatively careful manual scheduling; and "Optimal" represents the optimal results achieved by heuristic search described in this section.
The number in brackets indicates the number of instructions, and the number outside of the brackets indicates the clocks.

We can see that the optimal results are much more efficient than the results of the previous two methods, showing our brilliant pipeline performance of approximately 90\% efficiency.
This result implies that for our multiple-precision design, we achieve 90\% of the theoretical optimal floating-point performance.

\begin{table}
\footnotesize
\caption{Clocks of first pipeline based on different instruction scheduling methods.}
\label{tab:inssche}
\begin{tabular}{lcccc}
\toprule
Method & 128 Outer & 128 Inner & 256 Outer & 256 Inner \\
\midrule
Origin & 21(13) & 48(34) & 64(32) & 111(64) \\
Manual & 15(13) & 37(31) & 39(32) & 79(58) \\
Optimal & 15(13) & 33(31) & 35(32) & 65(58)\\
\bottomrule
\end{tabular}
\end{table}

\section{Evaluation}

\subsection{Performance Results}

This part demonstrates the performance results. Unless otherwise specified, we use 39,286 nodes (157,144 processes) of Sunway TaihuLight for experimentation, which is almost the entire capacity of the supercomputer (a total of 40,960 nodes).

\subsubsection{Time to Solution}
\begin{figure}
\centering
\includegraphics[width=0.4\textwidth]{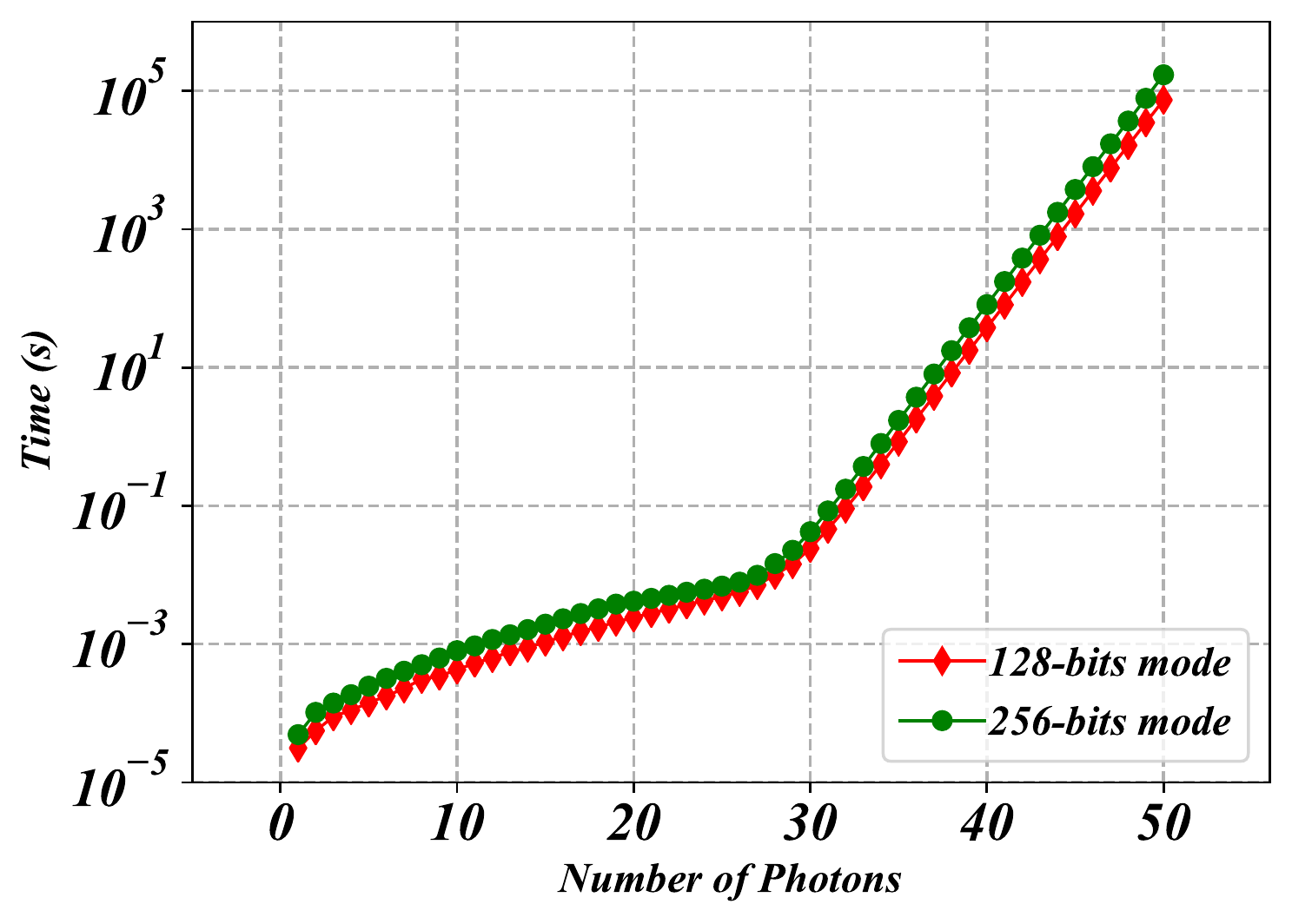}
\caption{Execution time for obtaining a single Torontonian. Two precision modes are tested with different number of photons on the entire Sunway TaihuLight supercomputer.}
\label{fig:time}
\end{figure}

Figure \ref{fig:time} shows the execution time of both precision modes based on a log-transformed distribution.
In the beginning, the execution time is nonlinear because the workload occupancy is insufficient.
Then, the time is almost linear in the end due to the high scalability.

To obtain a sample, it usually needs to calculate about 100 probabilities of the candidate samples using Markov Chain Monte Carlo (MCMC) sampling method \cite{neville2017classical}. 
To calculate one Torontonian probability for a 50-click photon detecting event, the execution time in our benchmark is 73,773s (about 20 hours) for 128-bit mode and 170,891s (about 2 days) for 256-bit mode, respectively.

\begin{figure}
\centering
\includegraphics[width=0.4\textwidth]{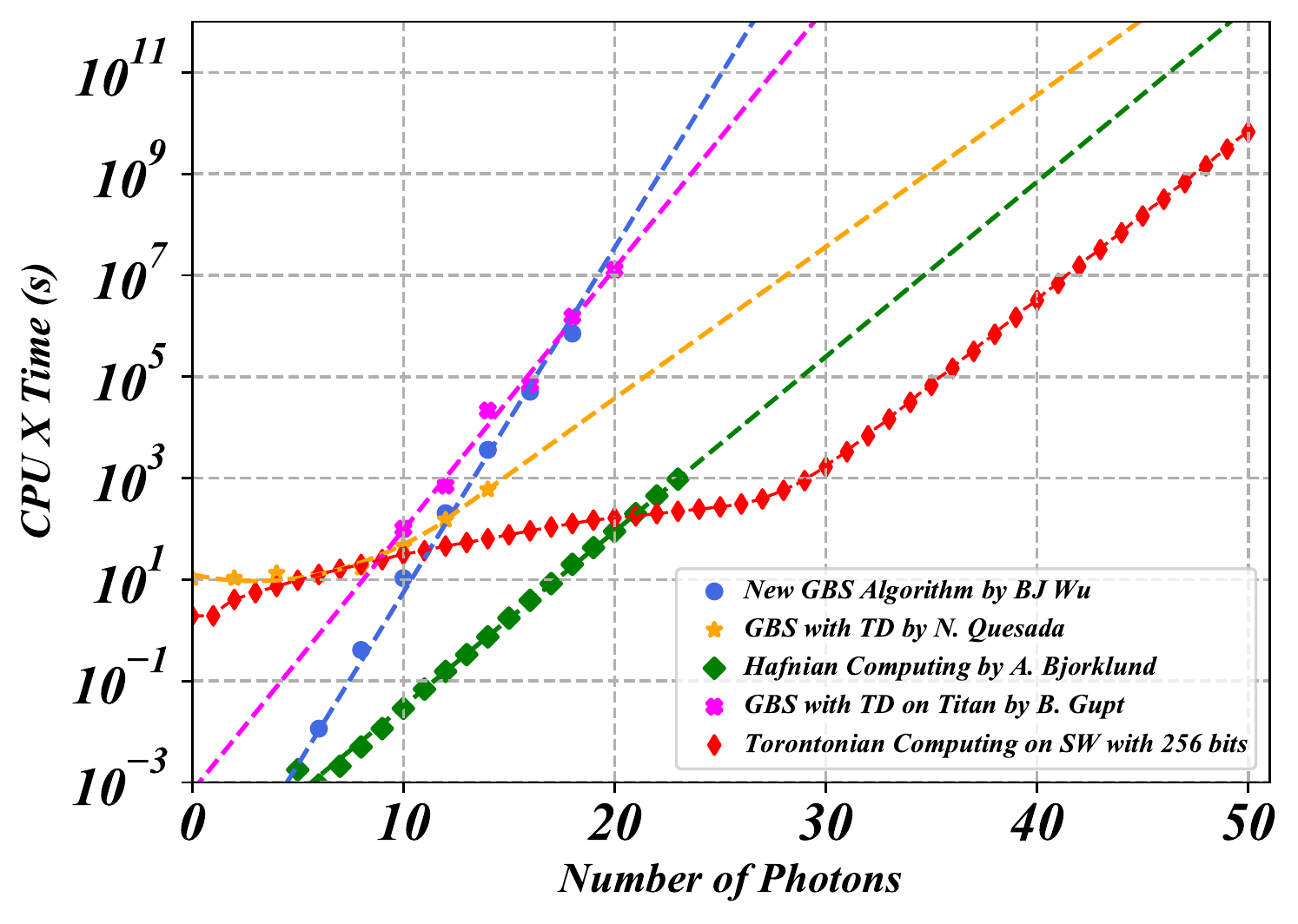}
\caption{$CPU\times Time$ results of various GBS classical benchmarks.
Red line indicates this work when using 256-bit precision mode.
Blue, yellow, green, and pink lines indicate the works of \cite{wu2020speedup}, \cite{quesada2019classical}, \cite{hafnian}, and \cite{gupt2020classical}, respectively.}
\label{fig:timecmp}
\end{figure}

Figure \ref{fig:timecmp} shows a comparison with other works that handle GBS. 
In the figure, we can see that except for our work (red line) and the work of \cite{hafnian} (green line), most works cannot process up to 50 photons.
Compared with work represented by the green line, our work is hundreds of times faster.
The time-to-solution results show the excellent performance of our method.

\subsubsection{FLOPS}
\label{subsec:flops}


For FLOPS counting, we count all arithmetic operations in the Torontonian function.
As a standard, solving a linear equation needs $2/3n^3+O(n^2)$ FLOPS with a size $n$ matrix \cite{top500}.
However, we need to make adjustments to the determinant calculation of the Torontonian function. 
First, as mentioned above, the input matrix is guaranteed to be an Hermitian positive definite complex matrix.
\textcolor{gray}{Thus, the cost of calculating a complex number should be considered, and half of the operations can be saved by Chomsky decomposition (or a modified Gaussian elimination).}
In addition, \textcolor{gray}{the matrix size $n$ is usually small in a Torontonian function}, so the term $O(n^2)$ should not be omitted.
Therefore, we count the operations of the linear solver accurately, as shown in Formula \ref{equ:linflops}.

\begin{equation}
\label{equ:linflops}
FLOPS = \frac{4}{3}n^3 + n^2 - \frac{4}{3}n
\end{equation}

Bringing this formula into the Torontonian function and using some combinatorial math skills, we obtain the total FLOPS:
\begin{equation}
\label{equ:tflops}
FLOPS = \sum_{i=1}^{N} \tbinom{N}{i} (\frac{32}{3}i^3 + 4i^2 - \frac{8}{3}i)
\end{equation}

\begin{figure}[!ht]
\centering
\begin{subfigure}{.35\textwidth}
\includegraphics[width=1\textwidth]{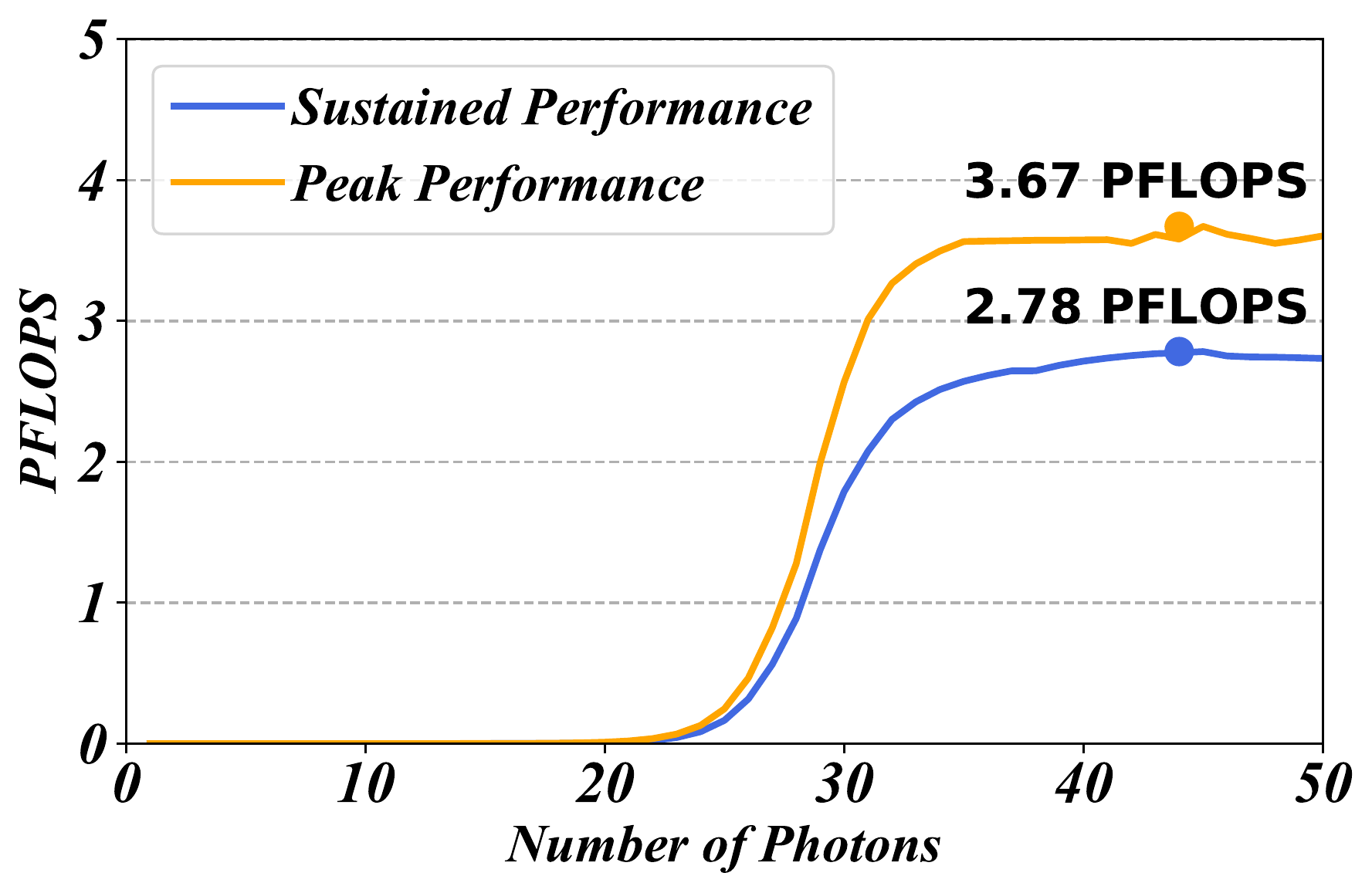}
\caption{}
\end{subfigure}
\begin{subfigure}{.35\textwidth}
\includegraphics[width=1\textwidth]{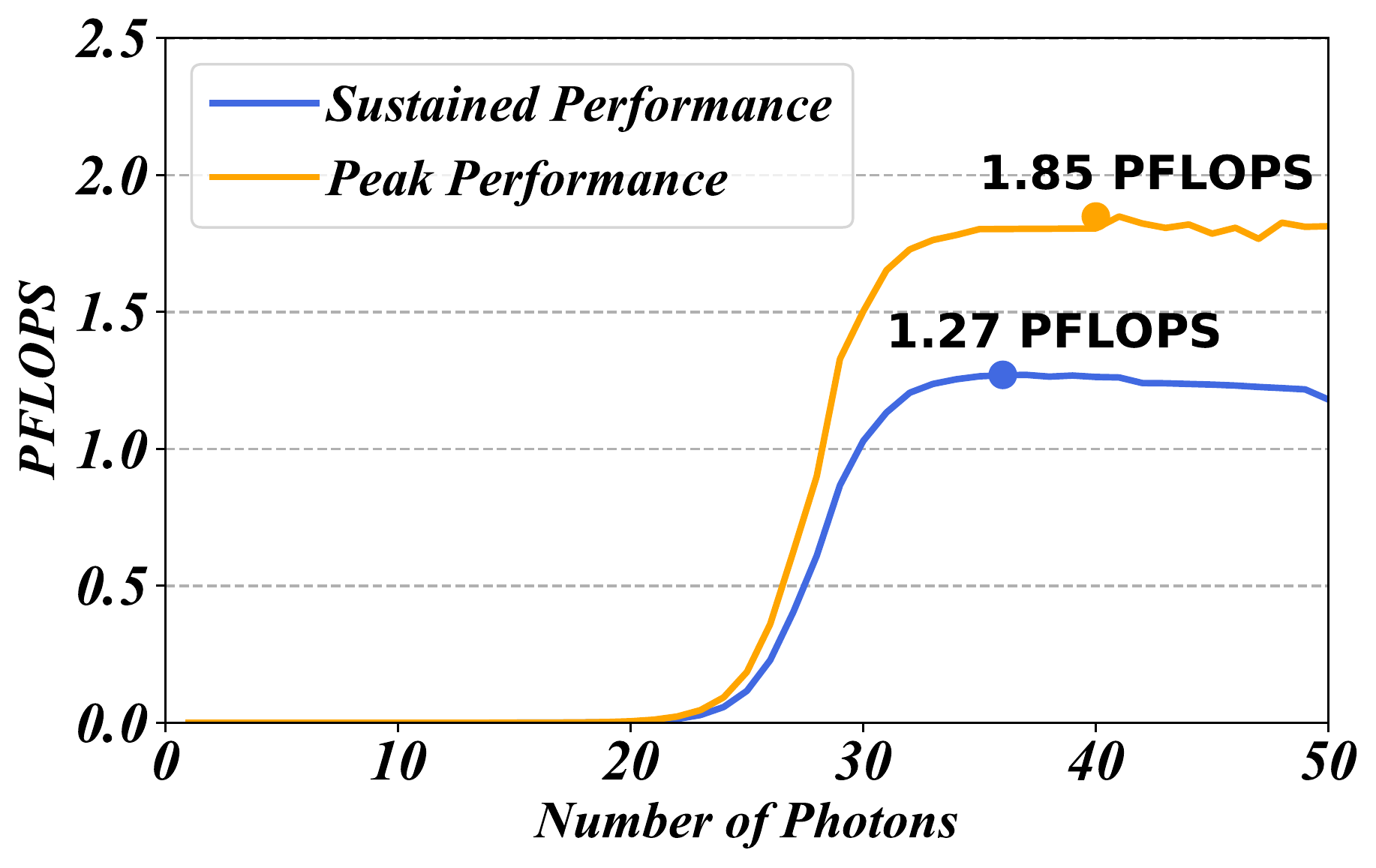}
\caption{}
\end{subfigure}
\caption{Peak and Sustained Performance of (a) 128-bit precision; (b) 256-bit precision.}
\label{fig:flops}
\end{figure}

Figure \ref{fig:flops} shows the performance results of two different precisions based on the final formula.
The highest sustained performance of 128-bit mode and 256-bit mode are $2.78$ PFLOPS when $N = 45$ and $1.27$ PFLOPS when $N = 37$;
the highest peak performance is $3.67$ PFLOPS when $N = 45$ and $1.85$ PFLOPS when $N = 41$, respectively.

The highest FLOPS did not occur at $N = 50$. 
This is because when N becomes larger, the LDM space requirement becomes larger.
Thus, as mentioned in Subsection \ref{subsec:storage}, a calculating matrix is scattered to more CPEs, which results in additional overhead (explained in Subsection \ref{subsec:kern}).
This issue is especially severe in 256-bit mode.

\subsubsection{Scalability}
\label{subsec:scala}
\begin{figure}[!ht]
\centering
\begin{subfigure}{.35\textwidth}
\includegraphics[width=1\textwidth]{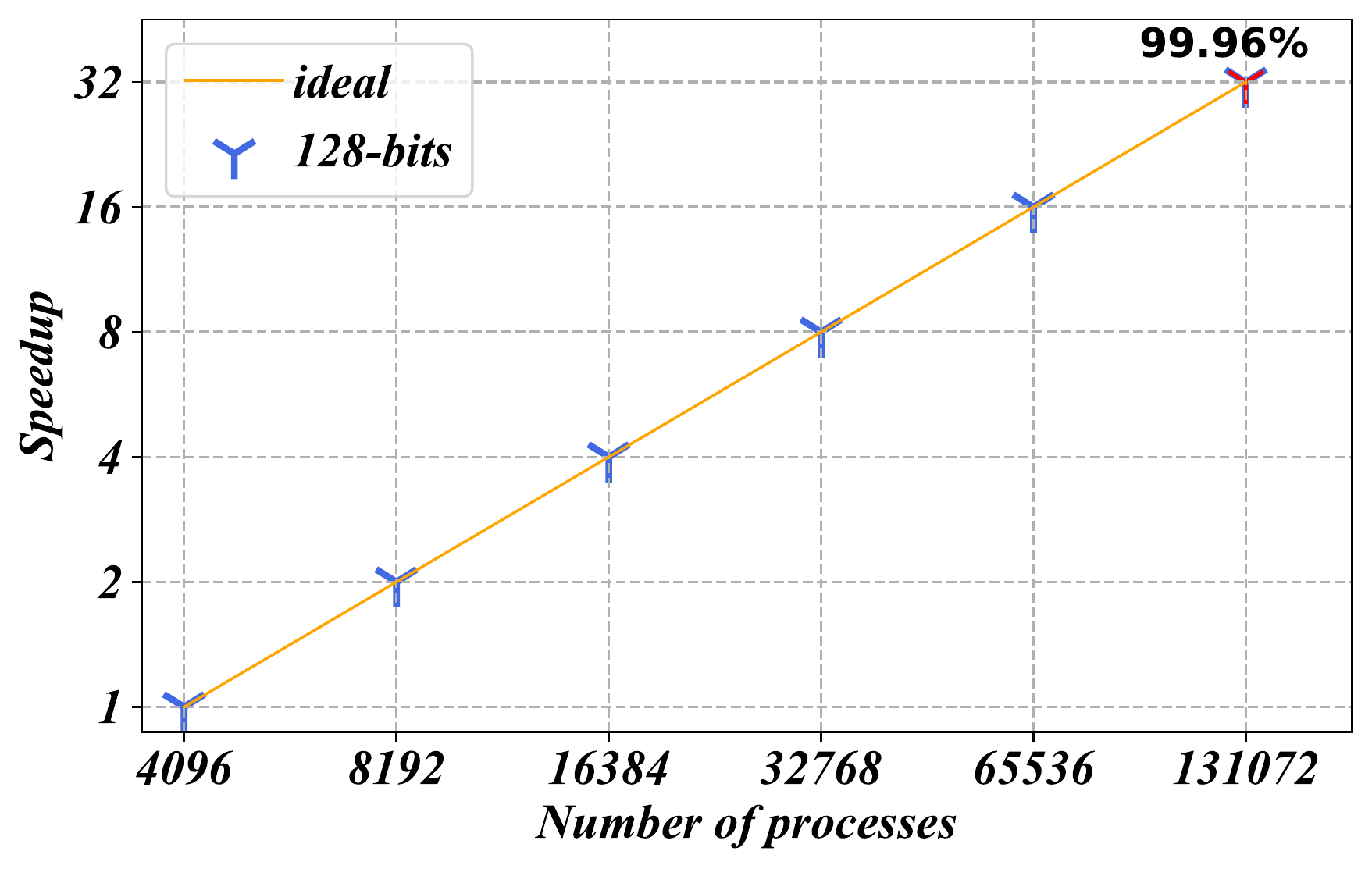}
\caption{}
\end{subfigure}
\begin{subfigure}{.35\textwidth}
\includegraphics[width=1\textwidth]{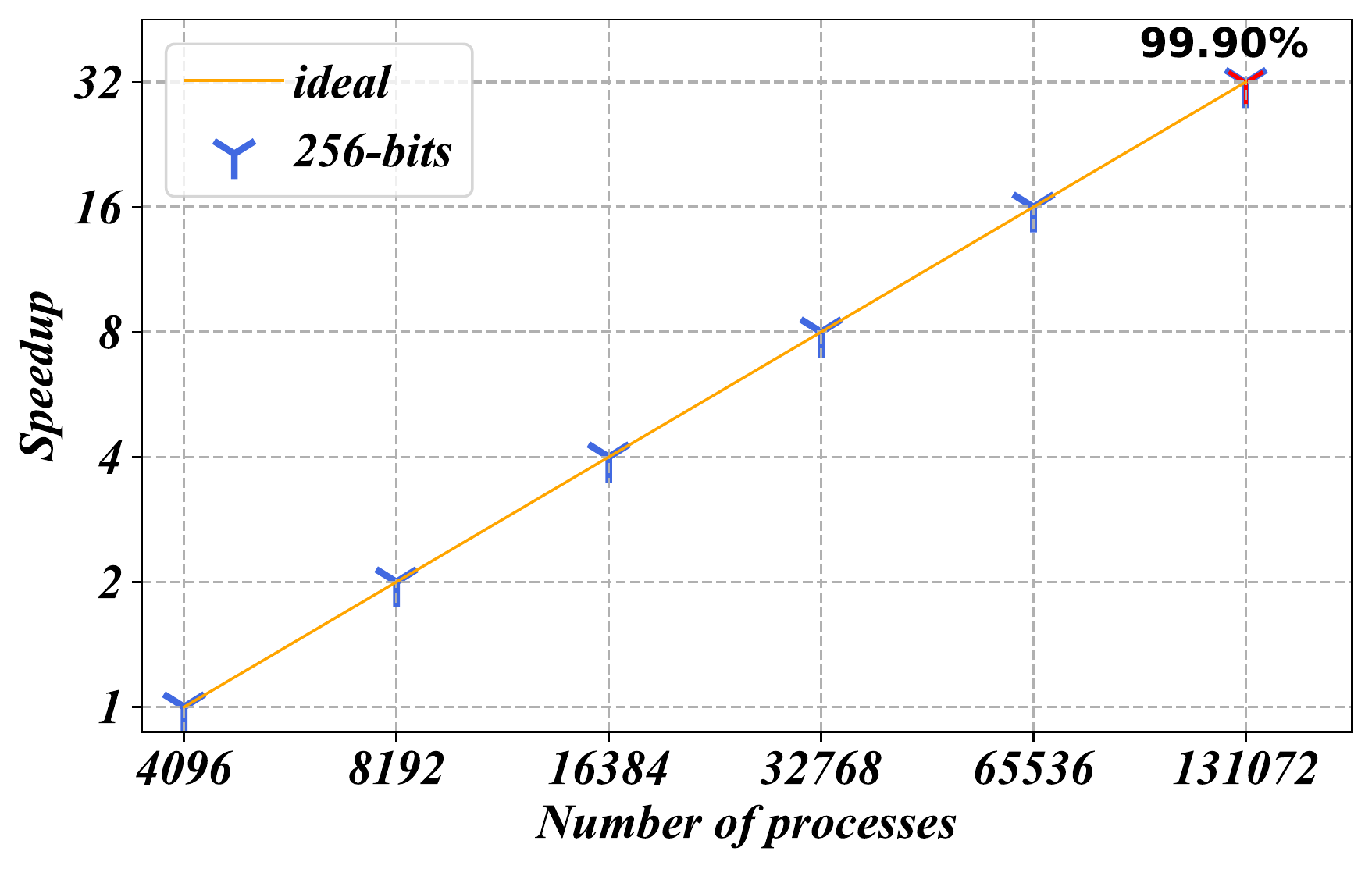}
\caption{}
\end{subfigure}
\caption{Strong scalability of (a) 128 bits; (b) 256 bits.}
\label{fig:scalability}
\end{figure}

For strong scalability, we choose the data scale $N = 36$ and use 4,096 to 131,072 processes.
Figure \ref{fig:scalability} shows the strong scalability results. 
As the figure shows, both precision modes achieve an almost linear strong scalability.
%

\subsubsection{Load Balance}

\begin{table}[!t]
\scriptsize
\caption{Load balancing results}
\label{tab:load}
\begin{tabular}{lcccc}
\toprule
Level & Average & Maxinum & Mininum \\
\midrule
Thread level & 55.6G & 55.6G & 55.6G \\
Process level & 55.6G & 55.6G & 55.6G \\
\bottomrule
\end{tabular}
\end{table}

This part reports the load balance performance with regard to the partition strategy described in Subsection \ref{subsec:partition}.
Table \ref{tab:load} shows the result in 128-bit precision mode when $N = 40$.
Regardless of the average, maximum, and minimum values, they are all 55.6G clocks.
Thus, as long as the data scale is large enough, our partition strategy can achieve an almost completely balanced load.

\subsubsection{Kernel Performance}
\label{subsec:kern}

\begin{figure}
\centering
\includegraphics[width=0.35\textwidth]{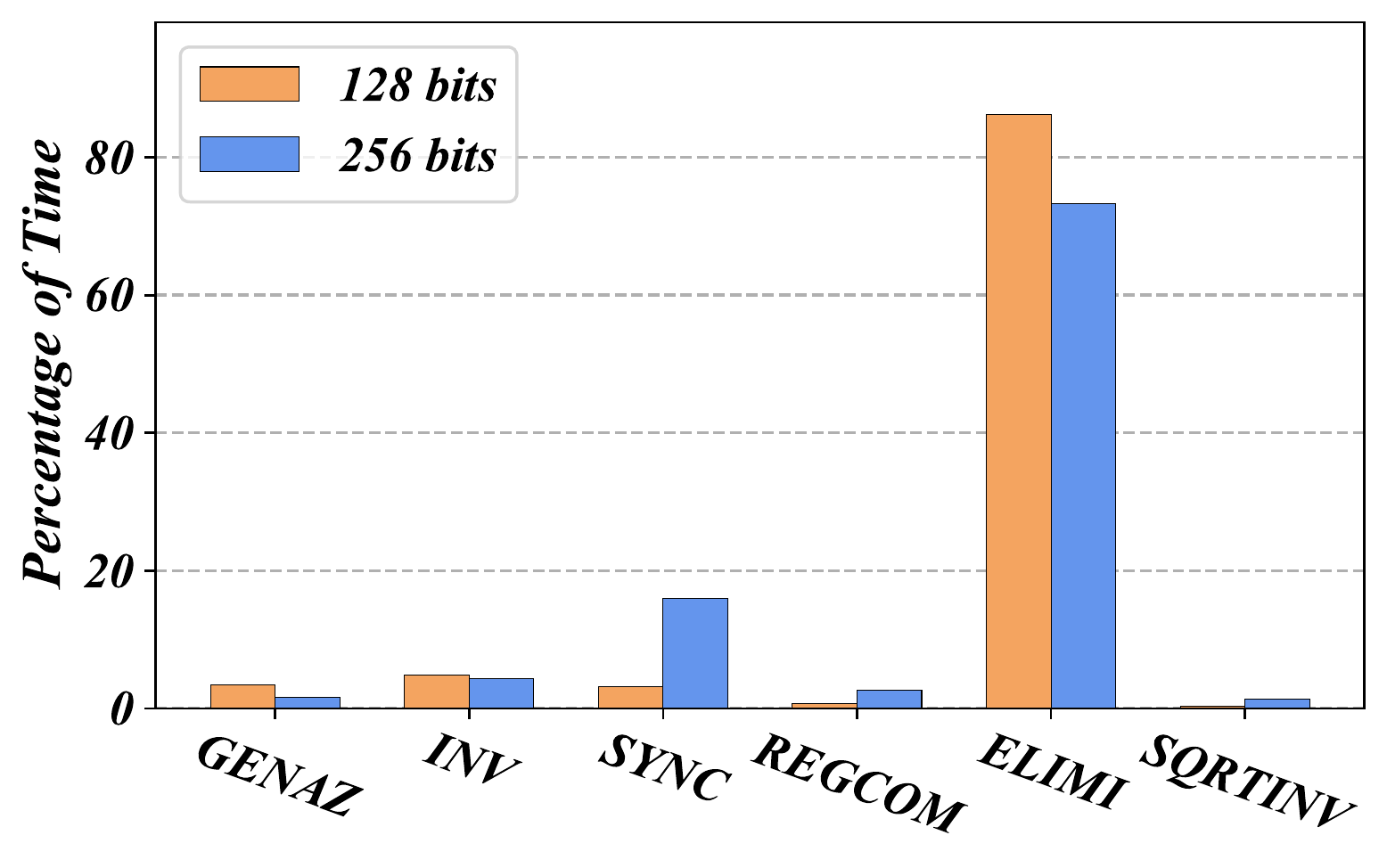}
\caption{\textcolor{gray}{Percentage of time cost of each kernel}}
\label{fig:kern}
\end{figure}

Figure \ref{fig:kern} shows the percentage of time cost of each kernel when $N = 50$. 
As shown in Algorithm \ref{alg:tor}, GENAZ corresponds to Line 6, which is used to generate the matrix $A_Z$ based on the current mask.
INV, SYNC, REGCOM, and ELIMI all correspond to Line 7, i.e., the function \textit{get\_determinant}. 
INV is related to the reciprocal calculation, while REGCOM is related to the register communication.

SYNC represents the time cost of synchronization with others' CPEs. 
This mainly occurs when several CPEs are working together to compute an identical matrix. Only one CPE needs to calculate the reciprocal of the pivot, and the other CPEs simply wait.
From Figure \ref{fig:kern} we can see the time cost of SYNC in 256 bits is much higher than that for 128 bits since the matrix is scattered to more CPEs in 256 bits.
This is the main factor that \textcolor{gray}{affects} performance when $N$ becomes larger.

ELIMI is related to the inner loop of the linear solver, which is composed of $O(2^NN^3)$ times of arithmetic and thus is the bottleneck that takes up $70\% \sim 90\%$ of the time.

Last, SQRTINV corresponds to Line 8, which computes the reciprocal square root.



\subsection{Comparisons with Other Architectures}

\begin{table}
\footnotesize
\caption{FLOPS compared with other architectures}
\label{tab:otherarch}
\begin{tabular}{lllllll}
\toprule
Hardware & Library & Kernel & Prec. & Peak. & Result \\
\midrule
SW26010 & - & $Tor(A)$ & 128 & 3.06T & \textcolor{red}{93.5G} \\
SW26010 & - & $Tor(A)$ & 256 & 3.06T & \textcolor{red}{46.5G} \\
E5-2680 v3 & DD & ADD & 106 & 0.96T & 15.9G \\
E5-2680 v3 & GMP & ADD & 128 & 0.96T & 0.9G \\
KP 920-4826 & DD & ADD & 106 & 1.00T & 0.89G \\
KP 920-4826 & GMP & ADD & 128 & 1.00T & 4.19G \\
Tesla C2050 & CAMPARY & GEMM & 106 & 0.51T & 14.8G \\
Tesla C2050 & CAMPARY & GEMM & 212 & 0.51T & 0.98G \\
\bottomrule
\end{tabular}
\end{table}

We use the FLOPS metric to compare the performance with other architectures.
For the x86 architecture (Xeon E5-2680 v3) and the ARM architecture (Kunpeng 920-4826 ARM v8), 
we directly run a series of additions without data dependence (called kernel "ADD") based on several efficient libraries such as DD and GMP.
For the GPU platform (Tesla C2050), we choose the CAMPARY library as our counterpart, which is one of the best multiple precision libraries in GPU. 
We directly use data supported by work \cite{joldes2017implementation} as the GPU results.
The work \cite{isupov2020multiple} proposes a new multiple precision algorithm that is faster than CAMPARY when requiring greater than 106 bits of precision.
However, in this work, all libraries are tested based on a single-precision version on the NVIDIA GTX architecture, which is much slower than the double-precision version.

Although the kernels used by each architecture are different, 
we guarantee that the Torontonian function used in our work is the most difficult one.

Table \ref{tab:otherarch} lists the FLOPS results among all architectures.
Only the GPU architecture is comparable to our algorithm on the Sunway architecture.
Since the Tesla C2050 is a relatively old version of the GPU, we also choose the Tesla V100 GPU (7.066 TFLOPS peak performance).
For this purpose, we convert the performance results based on the ratio of the peak performance of two GPUs.
The results of 106 bits and 212 bits are 204 GFLOPS and 13.5 GFLOPS, respectively.
Although the 106-bit mode is much better now, the 212-bit mode, which must be applied when the data scale is very large, is over three times slower than our algorithm.

Consequently, \textcolor{gray}{our algorithm based on the Sunway architecture has an advantage compared with other architectures.}

\section{Discussion}

In addition to introducing a state-of-the-art benchmark based on GBS with threshold detection, our work provides some new insights from the computing perspective.
Our partition strategy is applicable to other platforms.
Our heuristic search for instruction scheduling is suitable for other applications with a hand-write assembly kernel.
With minor modifications, our idea can also be applied to other platforms, especially for those platforms with a write-back stage structural hazard.

The multiple-precision fixed-point design is one of our major contributions.
Based on a \textcolor{gray}{architecture-specific} large integer instruction set, a set of fast operators is achieved to support the Torontonian function,
which includes addition, subtraction, multiplication, division, and reciprocal square root.
Our design can be easily applied to other scientific applications requiring high accuracy.
In addition to the Sunway architecture, any current or future machine with such a specific instruction set can benefit.
In our future work, we may increase and improve the operators to form an efficient and cross-platform fixed-point multiple-precision library based on the hardware instructions.

Last, our program has good portability by replacing our customized precision design with existing multiple-precision libraries under other platforms.

\section{Conclusion}
GBS with threshold detection is one of the most feasible ways to demonstrate quantum supremacy.
Based on the Sunway TaihuLight supercomputer, which \textcolor{gray}{contains} suitable architecture for this problem, we established a state-of-the-art classical benchmark waiting for its future quantum computing counterpart.

To simultaneously achieve almost optimal performance and sufficient accuracy, a series of methods including a partition strategy with excellent load balancing, optimal instruction scheduling based on a heuristic search, multiple-precision fixed-point design based on an architecture-specific instruction set, and adaptive precision mode selection according to upper- and lower-bound estimates was proposed and applied.

As a result, sustained performance of $2.78$ PFLOPS for 128-bit precision and $1.27$ PFLOPS for 256-bit precision was achieved with a proper $N$.
The largest run enabled us to obtain one Torontonian function of a $100\times100$ submatrix from 50-photon GBS within 20 hours with 128-bit precision and 2 days in 256-bit precision.

\bibliographystyle{unsrt}
\bibliography{quantum}

%
%
%
%
%
%
%




\end{document}